%% file: paper.tex
\documentclass[prl,superscriptaddress,twocolumn,showpacs,preprintnumbers,amsmath,amssymb]{revtex4}

\usepackage{subfigure}
\usepackage{epsfig}
\usepackage{dcolumn}
\usepackage{bm}
\usepackage{graphicx}
\usepackage{hyperref}

\def\simge{\mathrel{
   \rlap{\raise 0.511ex \hbox{$>$}}{\lower 0.511ex \hbox{$\sim$}}}}
\def\simle{\mathrel{
   \rlap{\raise 0.511ex \hbox{$<$}}{\lower 0.511ex \hbox{$\sim$}}}}

\newcommand {\GeV} {{\rm GeV}}

\newcommand{\MET}{\mbox{$E\kern-0.60em\raise0.10ex\hbox{/}_{T}$}}

\newcommand{\mH}{m_H}

\newcommand{\HWWstarlvlv}{H\rightarrow WW^{(\ast)}\rightarrow \ell\nu\ell\nu}
\newcommand{\HWWlvlv}{H\rightarrow WW\rightarrow \ell\nu\ell\nu}
\newcommand{\HWWlvqq}{H\rightarrow WW\rightarrow \ell\nu jj}

\begin{document}

\title{
{\vspace{-3.0 cm}\flushleft{\normalsize\normalfont{CERN-PH-EP-2011-130}}\flushright{\vspace{-0.86cm}\normalsize\normalfont{Submitted to Physical Review Letters}}\\[1.0cm]}
  Search for the Higgs boson in the $H\rightarrow
  WW\rightarrow\ell\nu jj$ decay channel \\ in $pp$ collisions at
  $\boldmath{\sqrt{s}}$ = 7\,TeV with the ATLAS detector}
\author{The ATLAS Collaboration}

\begin{abstract} 

A search for a Higgs boson has been performed in the $H\rightarrow
WW\rightarrow\ell\nu jj$ channel in 1.04 ${\rm fb}^{-1}$ of $pp$ collision data
at $\sqrt{s} = 7~{\rm TeV}$ recorded with the ATLAS detector at the 
Large Hadron Collider. No significant excess of events is observed over the
expected background and limits on the Higgs boson production cross section 
are derived for a Higgs boson mass in the range $240~{\rm GeV}<\mH<600$~GeV.
The best sensitivity is reached for $\mH = 400~\GeV$, where the 95\%
confidence level upper bound on the cross section for $H\rightarrow WW$
production is 3.1 pb, or 2.7 times the Standard Model prediction.
\end{abstract}


\maketitle

\normalsize

In the Standard Model (SM~\cite{prl_19_1264,np_22_579,sm_salam}), a
scalar field vacuum expectation value breaks the electroweak symmetry,
gives masses to the $W$ and $Z$
bosons~\cite{prl_13_321,prl_13_508,prl_13_585}, and manifests itself
directly as the so-called Higgs boson. A primary goal of the Large
Hadron Collider (LHC) is to test the SM mechanism of electroweak
symmetry breaking by searching for Higgs boson production in high
energy proton-proton collisions. Thanks in part to the large gluon
luminosity at LHC energies~\cite{Lai:2010vv,Sherstnev:2007nd}, the
Higgs boson is predominantly produced via gluon fusion ($gg\rightarrow
H$)~\cite{Anastasiou:2008tj,deFlorian:2009hc,Aglietti:2004nj,Actis:2008ug}
and to a lesser extent via vector boson fusion 
($qq\rightarrow
qqH$)~\cite{Bolzoni:2010xr,Ciccolini:2007jr,Ciccolini:2007ec}. Current
limits from direct searches at LEP and the Tevatron exclude Higgs boson masses $\mH <
114.4$~GeV~\cite{Barate:2003sz} and 156~GeV$< \mH <
177$~GeV~\cite{tevhiggs} at 95\% C.L.

For $m_H \simge 135~\GeV$, the dominant decay mode of the Higgs boson
is $H\rightarrow
WW^{(\ast)}$~\cite{Bredenstein:2007ec,LHCHiggsCrossSectionWorkingGroup:2011ti}.
The most sensitive Higgs boson search channel in the mass region around
$m_{H}=160~\GeV$ is the purely leptonic mode $\HWWstarlvlv$. For $m_H
\simge 200~\GeV$, the $\HWWlvqq$ channel, where one $W$ decays to a
pair of jets ($W \rightarrow jj$), also becomes important. The
advantage of $\HWWlvqq$ over $\HWWlvlv$ is the ability to fully
reconstruct the Higgs boson mass.

This Letter describes a search for a Higgs boson in the $\HWWlvqq$
channel using the ATLAS detector at the LHC, based on 1.04 fb$^{-1}$
of $pp$ collision data at a center-of-mass energy $\sqrt{s} = 7$ TeV collected during 2011. In
this analysis, the distribution of the $\ell\nu jj$ invariant mass
$m(\ell\nu jj)$, reconstructed using the charged-lepton neutrino
invariant mass constraint $m(\ell\nu) = m(W)$ and the requirement that
two of the jets in the event are consistent with a $W\rightarrow jj$
decay, is used to search for a Higgs boson signal. The results of a
similar search for $H\rightarrow WW\rightarrow\ell\nu jj$ based on 35
pb$^{-1}$ of data recorded during the 2010 LHC run were presented in
Ref.~\cite{Aad:2011qi}.

The present search, based on the measured shape of the $m(\ell\nu jj)$
distribution, is restricted to $m_{H} > 240$ GeV, in order to ensure a
smoothly varying non-resonant background, well clear of the effective
kinematic cutoff $m(\ell\nu jj) \sim 160$ GeV. For $m_H \simge
600~\GeV$, the jets from $W\rightarrow jj$ decay begin to overlap due
to the large $W$ boost, and the natural width of the Higgs boson
becomes large.  A detailed treatment of these issues is beyond the
scope of the present analysis. The best sensitivity in this analysis
is expected for $m_{H} \sim 400$ GeV.

The ATLAS detector~\cite{atlas} is a multipurpose particle physics
apparatus with forward-backward symmetric cylindrical geometry
covering the pseudorapidity range $|\eta|<2.5$ for track and
$|\eta|<4.9$ for jet measurements~\cite{footnote22}. The inner
tracking detector (ID) consists of a silicon pixel detector, a silicon
microstrip detector, and a transition radiation tracker. The ID is
surrounded by a thin superconducting solenoid providing a 2~T magnetic
field, and by a high-granularity liquid-argon (LAr) sampling
electromagnetic (EM) calorimeter. An iron-scintillator tile
calorimeter provides hadronic coverage in the central rapidity
range. The end-cap and forward regions are instrumented with LAr
calorimetry for both electromagnetic and hadronic measurements. The 
muon spectrometer surrounds the calorimeters and consists of three
large superconducting toroids, each with eight coils, a system of
precision tracking chambers, and detectors for triggering.

Detailed Monte Carlo (MC) studies of signal and backgrounds have been
carried out~\cite{ATLAS_simulation_paper}. The interaction with the
ATLAS detector is modeled with GEANT4~\cite{geant4} and the events are
processed using the same reconstruction that is used to perform the
reconstruction on data.  The effect of multiple $pp$ interactions in
the same bunch crossing (pile-up) at the high luminosities achieved by
the LHC in 2011 is modeled by superimposing, at the generation stage,
several simulated minimum-bias events on the simulated signal and
background events. MC samples were generated with different pile-up
levels and subsequently reweighted to match the pile-up conditions
observed in the data.

The data used in this analysis were recorded during periods when all
ATLAS sub-detectors were operating under nominal conditions. The
events were triggered by requiring the presence of an electron
candidate with transverse energy $E_{T}>20$~GeV or a muon candidate
with transverse momentum $p_{T}>18$~GeV.

Electron candidates are selected from clustered energy deposits in the
EM calorimeter with an associated track and are required to satisfy a
tight set of identification cuts~\cite{:2010yt} with an efficiency of
71 $\pm$ 1.6\% for electrons with $E_{T}>20$~GeV. While the energy
measurement is taken from the EM calorimeter, the pseudorapidity
$\eta$ and azimuthal angle $\phi$ are taken from the associated
track. The cluster is required to be inside the range $|\eta|<2.47$,
excluding $1.37<|\eta|<1.52$ and small calorimeter regions affected by
temporary operational problems. The track associated with the electron
candidate is required to point back to a reconstructed primary vertex
with a transverse impact parameter significance
$d_{0}/\sigma_{d_{0}}\le 10$ and an impact parameter along the
beam direction $z_{0}\le 10$~mm. Electrons are required to be
isolated: the sum of the transverse energies in cells inside a cone
$\Delta R<0.3$ around~\cite{footnote26} the cluster barycenter
(excluding the electron itself) must satisfy $\Sigma(E_{T}^{calo})<4$~GeV.

Muons are reconstructed by combining tracks in the inner detector and
muon spectrometer, with efficiency $92\pm 0.6$\% for muons with
$p_{T}>20$~GeV. Muons are required to pass basic quality cuts on the
number and type of hits in the inner detector. They must lie in the
range $|\eta|<2.4$, and satisfy the same impact parameter cuts as
electrons. They must also be isolated, with the sum of the transverse
momenta of all tracks in a cone $\Delta R<0.2$ around the muon
satisfying $\Sigma(p_{T}^{track})/p_{T}^{\mu}<0.1$.

Jets are reconstructed from topological clusters using the
anti-$k_{t}$ algorithm~\cite{AntiKt} with radius parameter
$R=0.4$. The reconstructed jets are calibrated using $E_{T}$ and
$\eta$ dependent correction factors based on MC simulation and
validated with data~\cite{jetEscaleNote}. They are required to have
$E_{T}>25$~GeV and $|\eta|<4.5$. Jets are considered $b$-tagged if
they contain a reconstructed displaced secondary vertex consistent
with a $b$-decay~\cite{btaggingCONF}. The operating point chosen for
this $b$-tag selection has an efficiency of 50\% for $b$-jets in
$t\overline{t}$ events in MC, and the mistag rate for non-$b$-jets has
been measured to be between 0.1\% and 2.0\%, depending on the $|\eta|$
and $p_{T}$ of the jet~\cite{btaggingCONF}. The event missing
transverse momentum \MET ~is reconstructed starting from topological
energy clusters in the calorimeters calibrated according to the type
of the object to which they are associated. The momenta of any muons
in the event are also taken into account in the $\MET$ measurement.

For this analysis, events are required to have at least one vertex
with at least three associated tracks with $p_{T}>400$~MeV. There must
be exactly one reconstructed lepton candidate (electron or muon) with
$p_{T}>30$~GeV. In order to ensure that this analysis is statistically
independent of the ATLAS $H\rightarrow ZZ\rightarrow\ell\ell\nu\nu$
analysis, events are vetoed if there is an additional lepton with
$p_{T}>20$~GeV, including electrons which only satisfy the looser
identification cuts used in the $H\rightarrow ZZ\rightarrow
\ell\ell\nu\nu$ analysis~\cite{ATLAS-CONF-2011-026}.

Events are required to have $\MET > 30$~GeV to account for an
unobserved neutrino from $W\rightarrow\ell\nu$ decay. There must be
exactly two jets ($H+0$ jet sample) or exactly three jets ($H+1$ jet
sample) with $E_T > 25$~GeV and $|\eta|<4.5$. The two jets with
invariant mass ($m_{jj}$) closest to the mass of the $W$ boson are
required to satisfy $71~\GeV < m_{jj} < 91~\GeV$. These two jets are
taken as the $W$ decay jets and are required to lie in the range
$|\eta|<2.8$, where the jet energy scale (JES) is best known (to
better than $\pm (4-8)$\% for $E_{T}>25$~GeV~\cite{jetEscaleNote}.)

After this event selection, the background is expected to be
dominated by $W$+jets production. Other important backgrounds are
$Z$+jets, multijets (MJ) from QCD processes, top quark, and diboson
($WW,WZ$, and $ZZ$) production. In order to further reject backgrounds
from top quark production, events are rejected if any of the jets is
$b$-tagged.

Although the MC is not used to model the background in the final fit
used to obtain limits, a combination of MC and data-driven methods is
used to better understand the background yields at this intermediate
stage. Backgrounds due to $W/Z$+jets, $t\overline{t}$, and diboson
production are modeled using the ALPGEN~\cite{alpgen},
MC@NLO~\cite{mcatnlo}, and HERWIG~\cite{herwig} generators,
respectively. A small contribution from $W/Z+\gamma$ events is
generated using MadEvent~\cite{Alwall:2007st}. The shape of the MJ
background is modeled using histograms derived from data samples
selected in an identical way to the $H\rightarrow WW\rightarrow\ell\nu
jj$ selection except that the electron identification requirements are
loosened and the isolation requirement on muons is inverted. In the
loosened selection, electrons satisfying the complete set of
identification criteria are not included. Expected contributions from
non-QCD processes to the MJ shape histogram are subtracted using MC
predictions.

To normalize the MJ shape histogram, the loose lepton control sample
selection is further relaxed by removing the $\MET$ cut to construct a
shape template for the $\MET$ distribution for the MJ background. The
normalizations of this MJ template and the corresponding template for
$W/Z$+jets taken from MC are fit to the observed $\MET$ distribution,
and the resulting scale factors are then used to normalize the MJ and
$W/Z$+jets processes in comparisons between data and
expectations. Both the gluon fusion and the vector boson fusion signal
production processes are simulated using the
POWHEG~\cite{powheg,Nason:2009ai} event generator interfaced to
PYTHIA~\cite{pythia}, normalizing to the NNLO cross
sections~\cite{LHCHiggsCrossSectionWorkingGroup:2011ti} shown in
Table~\ref{signalCrossSectionTable}.
\begin{table}[tb]
\begin{center}
\caption{Cross section for Standard Model Higgs boson production and
  the branching ratio (BR) for $H\rightarrow WW\rightarrow\ell\nu jj$
  ($\ell=e/\mu$) as a function of mass~\cite{LHCHiggsCrossSectionWorkingGroup:2011ti}.}
\begin{tabular}{ c | c c | c }
\hline
\hline
$\mH$[GeV]					& $\sigma(gg\rightarrow H)$ [pb]	& $\sigma(qq\rightarrow H)$ [pb] & BR($H\rightarrow\ell^{\pm}\nu jj$)\\
\hline
300\rule[-1mm]{0mm}{4.7mm}			& 2.4$\pm 0.4$				& 0.30$^{+0.014}_{-0.008}$       & 0.202\\    
400\rule[-1mm]{0mm}{4.7mm}			& 2.0$^{+0.31}_{-0.34}$			& 0.162$^{+0.010}_{-0.005}$      & 0.170\\   
500\rule[-1mm]{0mm}{4.7mm}			& 0.85$\pm 0.15$			& 0.095$^{+0.0068}_{-0.0032}$    & 0.160\\  
600\rule[-2mm]{0mm}{5.7mm}			& 0.33$^{+0.063}_{-0.058}$		& 0.058$^{+0.005}_{-0.002}$       & 0.164\\ 

\hline
\hline
\end{tabular}
\label{signalCrossSectionTable}
\end{center}
\end{table}

In order to reconstruct the invariant mass $m(\ell\nu jj)$ of the
$WW$ system, the mass constraint $m(\ell\nu)=m(W)$ is used, where
the neutrino transverse momentum $p_{T}^{\nu}$ is taken from the event
$\MET$. This equation can have real or complex solutions. In the case
of complex solutions, the event is rejected. This requirement rejects
45\% of background events in both data and MC, but only 36\% of MC
signal events with $m_{H}=400$~\GeV. In the case of two real
solutions, the solution with smaller neutrino longitudinal momentum
$|p_{z}^{\nu}|$ is taken, based on simulation
studies. Table~\ref{ggHcutflowTable} shows the observed and expected
numbers of events for signal and background after this full selection.
\begin{table*}
\caption{
Expected and observed numbers of events for an integrated luminosity
of 1.04~fb$^{-1}$ after all selection cuts (including the requirement
that $m(\ell\nu)=m(W)$ has a real solution) for the signal and the
main backgrounds. For the $W/Z$+jets and MJ backgrounds, the
uncertainties are taken from the fit to the $\MET$ distribution used
to normalize these backgrounds. For signal, top and diboson, the
quoted uncertainties are JES ($\pm$17\%), jet energy resolution (8.6\%), cross section ($\pm$10\% for
both top and diboson, and $\pm$ 19.4\% for signal), and luminosity
($\pm$3.7\%), added in quadrature; the total errors in the rightmost
column for these processes are the linear sum of the errors for the
individual channels since these sources of systematic uncertainty are
correlated across channels. Statistical errors are small compared to
these uncertainties.
}
\begin{ruledtabular}
\begin{tabular}{ c | r r  r r | r }
                           & H($e\nu jj$) + 0j  &  H($\mu\nu jj$) + 0j  &  H($e\nu jj$) + 1j  &  H($\mu\nu jj$) + 1j  &  H + 0j or 1j \\
\hline
$W/Z$+jets                      &    5580 $\pm$ 150       &         6690 $\pm$ 430    &     3610 $\pm$ 140        &    3940 $\pm$ 360         &   19820 $\pm$ 600 \\
Multi-jet                     &      592 $\pm$  16       &          222  $\pm$  15    &    420 $\pm$  16        &     195 $\pm$ 17          &    1430 $\pm$ 30 \\
Top                           &      87 $\pm$  17       &           87 $\pm$ 18     &      271 $\pm$ 54         &     280 $\pm$ 56         &    730 $\pm$ 150 \\
Dibosons                      &      226 $\pm$  45       &          222 $\pm$ 45     &      89 $\pm$ 18         &     112 $\pm$ 23          &     650 $\pm$ 130 \\
\hline
Expected Background           &    6490 $\pm$ 160       &         7220 $\pm$ 440    &     4390 $\pm$ 150        &    4530 $\pm$ 370         &   22630 $\pm$ 640 \\
\hline
Data                          &    6446                &           7201           &        4134              &       4380                &      22161  \\
\hline
Expected Signal ($m_H = 400$ GeV)  &     11 $\pm$ 3.0   &        9 $\pm$ 2.4       &       13 $\pm$ 3.5       &       10 $\pm$ 2.7        &      43 $\pm$ 12 \\
\end{tabular}
\end{ruledtabular}
\label{ggHcutflowTable}
\end{table*}

\begin{figure}[tb]
\begin{center}
\includegraphics[angle=0,width=0.43\textwidth]{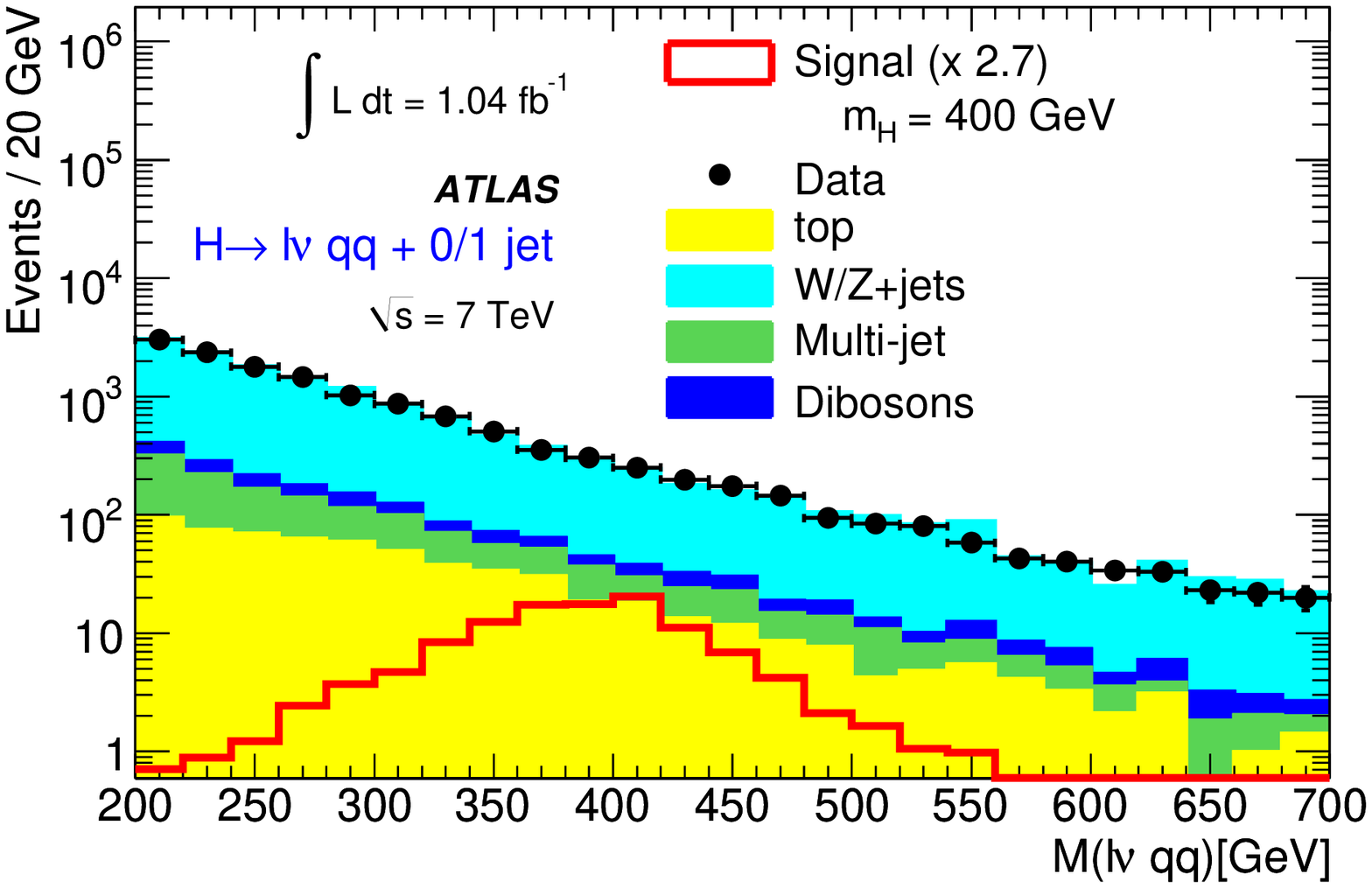} 
\includegraphics[angle=0,width=0.43\textwidth]{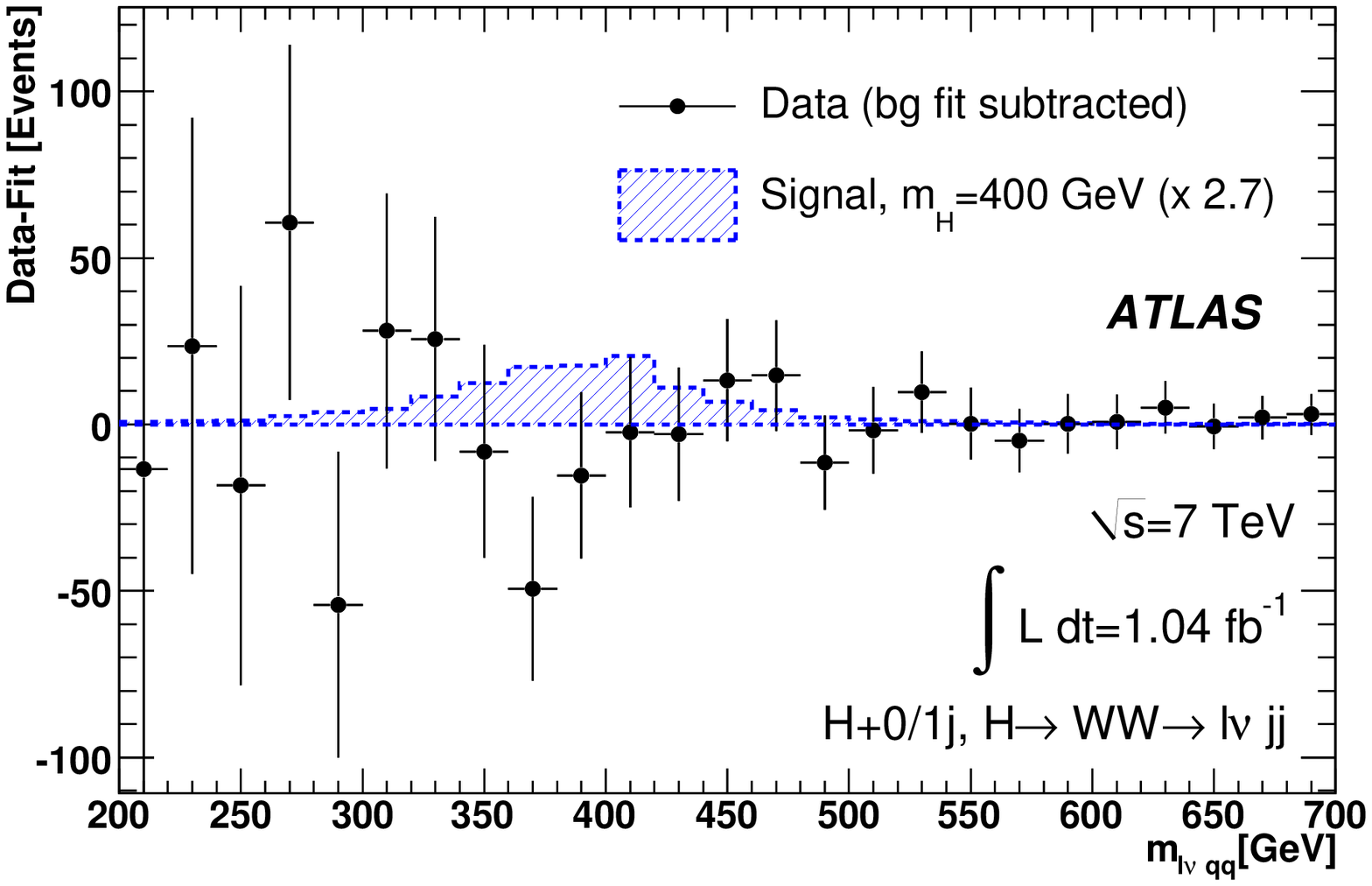}
\caption{
Top: the reconstructed invariant mass $m(\ell\nu jj)$ in the data
summed over lepton flavor and jet multiplicity. The expected
backgrounds are also shown. Bottom: the difference between data and
the fitted non-resonant background. The expected contribution from
Higgs boson decays for $\mH = 400$ GeV in the SM is also shown,
multiplied by a factor of 2.7.
}
\label{mlnuqq01j_plots}
\end{center}
\end{figure}

\begin{figure}[tb]
\begin{center}
\includegraphics[angle=0,width=0.43\textwidth]{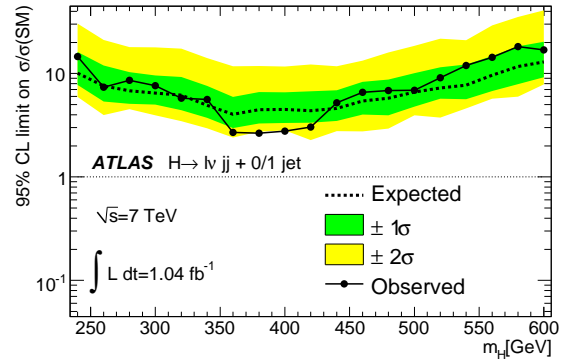}
\caption{
The expected and observed 95\% confidence level upper limits on the
Higgs boson production cross section divided by the SM prediction for
an integrated luminosity of 1.04 fb$^{-1}$. For any hypothesized Higgs
boson mass, the background contribution used in the calculation of
this limit is obtained from a fit to the $m(\ell\nu jj)$
distribution. The green and yellow bands show the $\pm 1\sigma$ and
$\pm 2\sigma$ uncertainties on the expected limit.
}
\label{limit_plot}
\end{center}
\end{figure}

Figure~\ref{mlnuqq01j_plots} (top) shows the $m(\ell\nu jj)$
distribution for this final sample. The expected signal for
$\mH = 400~\GeV$ is also shown, scaled up by a factor of 2.7. 
The $m(l \nu jj)$ resolution is $(7.5 \pm 0.6$)\% at $\mH=400$ GeV,
depending mostly on the jet energy resolution as checked in data versus MC
by various jet-balance techniques \cite{CONF-2010-054}, and shows a
$1/\sqrt{\mH}$ dependence over the range of this analysis. Limits
are set using a maximum likelihood fit to the shape of the observed
$m(\ell\nu jj)$ distribution in the range $200<m(\ell\nu
jj)<2000$~GeV. The non-resonant background in this fit is modeled by
the sum of two exponential functions. The normalization and slope of
each exponential are unconstrained parameters in the fit. The
double-exponential form for the total background is well justified by
fits to the $m(\ell\nu jj)$ distributions obtained by selecting events
with $m_{jj}$ just below ($50<m_{jj}<60$~GeV) or just above
($100<m_{jj}<110$~GeV) the $W$ peak, respectively.  As a consistency
check, the background parametrization was altered to use three
exponentials and the shift in signal yield as compared to the nominal
background shape was found to be small as compared to other
uncertainties. The $m(\ell\nu jj)$ distribution for the expected
signal at each hypothesized $m_H$ is modeled using the signal MC
samples. 

The fit includes nuisance parameters which account for the uncertainty
in the efficiency of the electron, muon, and jet reconstruction. The
electron and muon efficiencies are varied within their uncertainties,
leading to an uncertainty in the signal efficiency of $\pm$1.6\% and
$\pm$0.6\%, for electrons and muons respectively. Varying the jet
energy scale within its uncertainties yields a corresponding
uncertainty of $\pm$17\% in the expected signal, and smearing the jet
energies within the uncertainty on their resolutions results in a
signal uncertainty of $\pm$8.6\%. The limits also take into account a
$\pm$3.7\% uncertainty on the luminosity
determination~\cite{ATLAS-CONF-2011-116} and a $\pm$19.4\% uncertainty
on the predicted cross
section~\cite{LHCHiggsCrossSectionWorkingGroup:2011ti}, taken to be
independent of mass. The off-shell effects and interference between
the signal and backgrounds, which are discussed in
Refs.~\cite{LHCHiggsCrossSectionWorkingGroup:2011ti,Anastasiou:2011pi,Campbell:2011cu},
have been neglected. A conservative estimate of this uncertainty would
be 150$\times m_{H}^{3}$ ($m_{H}$ in TeV), where the $m_{H}^{3}$ form
is motivated by the scaling of the Higgs width with $m_{H}$ and the
normalization factor is chosen to give $\sim$30\% at $m_{H}=600$~GeV,
based on Figure~6 of Ref.~\cite{Campbell:2011cu}. If this were
included, it would increase the total systematic error by less than
$6$\% for $m_{H}\le 500$~GeV, and as much as 15\% for $m_{H}=600$~GeV
where the limit would be increased by $\sim 18$\%.

Figure~\ref{mlnuqq01j_plots} (bottom) shows the difference between the
$m(\ell\nu jj)$ distribution in data and the fitted background. There
is no indication of any significant excess. Limits are extracted using
the Profile Likelihood~\cite{asymptotics} as a test statistic and
following the $CL_{s}$ procedure described in Ref.~\cite{cls}. 

Figure~\ref{limit_plot} shows the 95\% CL upper bound on the cross-section
times branching ratio for Higgs production in units of the Standard Model 
prediction, $\sigma\times
BR_{H\rightarrow WW}/(\sigma\times BR_{H\rightarrow WW})_{\rm SM}$, as
a function of $\mH$. The observed cross section limit for $\mH =
400~\GeV$ is 3.1 pb, or 2.7 times the SM prediction, while the
corresponding expected limits are 5.2 pb or 4.5 times the SM
expectation. In the SM with an additional heavy fourth
generation~\cite{Kribs:2007nz,Unel:SM4}, the gluon fusion mechanism
for production of a Higgs boson is expected to be substantially
enhanced. Within the four generation context, a Higgs boson is
excluded at 95\% CL by the present data over the range $m_{H}=310-430$
GeV.

\section{Acknowledgements}

We thank CERN for the very successful operation of the LHC, as well as the
support staff from our institutions without whom ATLAS could not be
operated efficiently.

We acknowledge the support of ANPCyT, Argentina; YerPhI, Armenia; ARC,
Australia; BMWF, Austria; ANAS, Azerbaijan; SSTC, Belarus; CNPq and FAPESP,
Brazil; NSERC, NRC and CFI, Canada; CERN; CONICYT, Chile; CAS, MOST and
NSFC, China; COLCIENCIAS, Colombia; MSMT CR, MPO CR and VSC CR, Czech
Republic; DNRF, DNSRC and Lundbeck Foundation, Denmark; ARTEMIS, European
Union; IN2P3-CNRS, CEA-DSM/IRFU, France; GNAS, Georgia; BMBF, DFG, HGF, MPG
and AvH Foundation, Germany; GSRT, Greece; ISF, MINERVA, GIF, DIP and
Benoziyo Center, Israel; INFN, Italy; MEXT and JSPS, Japan; CNRST, Morocco;
FOM and NWO, Netherlands; RCN, Norway; MNiSW, Poland; GRICES and FCT,
Portugal; MERYS (MECTS), Romania; MES of Russia and ROSATOM, Russian
Federation; JINR; MSTD, Serbia; MSSR, Slovakia; ARRS and MVZT, Slovenia;
DST/NRF, South Africa; MICINN, Spain; SRC and Wallenberg Foundation,
Sweden; SER, SNSF and Cantons of Bern and Geneva, Switzerland; NSC, Taiwan;
TAEK, Turkey; STFC, the Royal Society and Leverhulme Trust, United Kingdom;
DOE and NSF, United States of America.

The crucial computing support from all WLCG partners is acknowledged
gratefully, in particular from CERN and the ATLAS Tier-1 facilities at
TRIUMF (Canada), NDGF (Denmark, Norway, Sweden), CC-IN2P3 (France),
KIT/GridKA (Germany), INFN-CNAF (Italy), NL-T1 (Netherlands), PIC (Spain),
ASGC (Taiwan), RAL (UK) and BNL (USA) and in the Tier-2 facilities
worldwide.

\bibliography{paper,Higgs_paper}

\onecolumngrid
\clearpage
\input{atlas_authlist}

\end{document}

%% file: atlas_authlist.tex
\begin{flushleft}
{\Large The ATLAS Collaboration}

\bigskip

G.~Aad$^{\rm 48}$,
B.~Abbott$^{\rm 111}$,
J.~Abdallah$^{\rm 11}$,
A.A.~Abdelalim$^{\rm 49}$,
A.~Abdesselam$^{\rm 118}$,
O.~Abdinov$^{\rm 10}$,
B.~Abi$^{\rm 112}$,
M.~Abolins$^{\rm 88}$,
H.~Abramowicz$^{\rm 153}$,
H.~Abreu$^{\rm 115}$
% ...
\bigskip

% ...
$^{10}$ Institute of Physics, Azerbaijan Academy of Sciences, Baku, Azerbaijan\\
$^{11}$ Institut de F\'isica d'Altes Energies and Departament de F\'isica de la Universitat Aut\`onoma  de Barcelona and ICREA, Barcelona, Spain\\
% ...
$^{48}$ Fakult\"{a}t f\"{u}r Mathematik und Physik, Albert-Ludwigs-Universit\"{a}t, Freiburg i.Br., Germany\\
$^{49}$ Section de Physique, Universit\'e de Gen\`eve, Geneva, Switzerland\\
% ...
$^{88}$ Department of Physics and Astronomy, Michigan State University, East Lansing MI, United States of America\\
% ...
$^{111}$ Homer L. Dodge Department of Physics and Astronomy, University of Oklahoma, Norman OK, United States of America\\
$^{112}$ Department of Physics, Oklahoma State University, Stillwater OK, United States of America\\
% ...
$^{115}$ LAL, Univ. Paris-Sud and CNRS/IN2P3, Orsay, France\\
% ...
$^{118}$ Department of Physics, Oxford University, Oxford, United Kingdom\\
% ...
$^{153}$ Raymond and Beverly Sackler School of Physics and Astronomy, Tel Aviv University, Tel Aviv, Israel\\
% ...
\end{flushleft}


%% file: paper.bbl
\begin{thebibliography}{43}
\bibitem{footnote22}{ATLAS uses a right-handed coordinate system with its origin at the nominal interaction point (IP) in the center of the detector and the $z$-axis coinciding with the axis of the beam pipe. The $x$-axis points from the IP to the center of the LHC ring, and the $y$-axis points upward. Cylindrical coordinates $(r, \phi )$ are used in the transverse plane, $\phi $ being the azimuthal angle around the beam pipe. The pseudorapidity is defined in terms of the polar angle $\theta $ as $\eta = -\protect \qopname  \relax o{ln}\protect \qopname  \relax o{tan}(\theta /2)$.}
\bibitem{footnote26}{$\Delta R=\protect \sqrt  {(\Delta \phi )^{2}+(\Delta \eta )^{2}}$, where $\Delta \phi$ and $\Delta \eta$ are the relevant separations  in $\phi$ and $\eta$, respectively.}
\end{thebibliography}
